# FPGA acceleration of Model Predictive Control for Iter Plasma current and shape control

Samo Gerkšič, Boštjan Pregelj, and Matija Perne

*Abstract*— A faster implementation of the Quadratic Programming (QP) solver used in the Model Predictive Control scheme for Iter Plasma current and shape control was developed for Xilinx Field-Programmable Gate Array (FPGA) platforms using a high-level synthesis approach. The QP solver is based on the dual Fast Gradient Method (dFGM). The dFGM is essentially an iterative algorithm, where matrix-vector arithmetic operations within the main iteration loop may be parallelized. This type of parallelism is not well-suited to standard multi-core processors because the number of operations to be spread among processing threads is relatively small considering the time-scale of thread scheduling. The FPGA implementation avoids this issue, but it requires specific techniques of code optimization in order to achieve faster solver execution.

## I. Introduction

MODEL Predictive Control (MPC) is an advanced control technique, which is already well established in the process industry for control of multivariable systems, particularly where advanced handling of constraints on process signals is important. In the recent years, its application for plasma magnetic control in magnetically-confined tokamak fusion reactors has been actively considered. In particular, an MPC controller for Iter plasma current and shape control (PCSC) has been developed [1]. In this controller, a dual Fast Gradient Method (dFGM) solver based on the QPgen library [2] is used for the on-line Quadratic Programming (QP) optimization that must be carried out at each sampling time. With MPC complexity reduction approaches and some optimization of the C code, the required solution accuracy is reached with computation times around 3 ms using a standard Intel central processing unit (CPU) in a personal computer in a single-threaded implementation. This implementation is considered sufficiently fast for Iter PCSC application, where a sampling time of 100 ms is considered appropriate for MPC control. Nevertheless, faster execution is desired for possible experimental evaluation of MPC control on dynamically faster medium-sized tokamaks.

The clock frequency, which is the main factor affecting single-threaded performance, has not increased considerably in the recent generations of CPUs. The computing capability is mainly improved by adding more computing cores. These may speed up execution of algorithms that can be parallelized efficiently. Unfortunately, this does not appear to be the case with the dFGM algorithm with relatively small problem dimensions in [1]. In dFGM, most of the computation time is spent in 500 cycles of the main iteration loop, which must be carried out sequentially. Parallelisation is possible only for matrix-vector arithmetic operations within each iteration. Spreading the computation to 4 CPU cores, each of the four threads would require around 1 μs per iteration. Unfortunately, the thread scheduler of the CPU operates at a time scale of around 10 μs, which makes such parallelisation highly inefficient.

The FPGA platform is considered promising for solver implementation because it naturally allows efficient micro-parallelism. However, the success of dFGM solver implementation was not considered certain because of restricted chip resources and relatively low clock frequencies compared to CPUs. In addition, specific programming skills are required for the implementation.

Different approaches to FPGA implementation of the dFGM solver are possible:

• Manual coding in hardware description languages (HDLs), such as VHDL and Verilog, considered the most flexible, but requires manual transcoding.

• A High Level Synthesis (HLS) approach where the HDL code is automatically generated from C code requires less time for the implementation and reduces the possibility of transcoding errors.

• The HLS approach is also possible from the development environment of Matlab code or Simulink schemes using HDL Coder of The Mathworks, Inc. The dFGM algorithm is also available as Matlab code for debugging purposes, however the C-code implementation is considered more efficient. HLS conversion of Simulink control schemes has the advantage of allowing direct implementation of control schemes without manual transcoding and simplifies documentation; however, for algorithms like dFGM coding in Simulink is not convenient and does not produce comprehensible documentation.

This paper was submitted 15.6.2018. Research supported by Slovenian Research Agency (P2-0001). This work has been carried out within the framework of the EUROfusion Consortium and has received funding from the Euratom research and training programme 2014-2018 under grant agreement No 633053. The views and opinions expressed herein do not necessarily reflect those of the European Commission.

Samo Gerkšič, Boštjan Pregelj, and Matija Perne are with Department of Systems and Control, Jožef Stefan Institute, Ljubljana, Slovenia (e-mail: samo.gerksic@ijs.si, bostjan.pregelj@ijs.si, Matija.Perne@ijs.si).

In this work, the C-code HLS approach is used for the FPGA implementation of the dFGM solver. The solver C code is manually adapted to a simpler C-language form acceptable for Xilinx Vivado HLS design suite. Basic requirements for the input C code are static declaration of variables' memory and direct implementation (in-lining) of certain functions. Vivado HLS then automatically converts the adapted C code to the VHDL FPGA programming language. Regarding the target FPGA hardware, the moderately-priced Xilinx Zedboard and the more capable Xilinx ZC706 are considered.

## II. RESULTS OUTLINE

With the initial automatic conversion from the original C code, the achieved results were correct but not useful, because the code structure prevents automated application of conversion optimization routines. With relatively modest usage of the FPGA resources the execution time with single flow-point numerical precision is 45 ms, which is much longer than the original CPU implementation (3 ms).

However, Vivado HLS supports multiple conversion directives ("pragmas") that enable optimized VHDL implementation, e.g. pipeline, loop unroll and merge, array partition. After a series of manual modifications of the code, the computation time is reduced to 1 ms using the Xilinx ZC706.

In the outline, the code of the dFGM solver consists of three parts: the prologue, the main iteration loop, and the epilogue. It is particularly important to optimize the execution of the main loop, because it needs to cycle 500 times in order to reach the required precision.

One iteration of the loop is a sequence of two matrix-vector multiplications, amounting to about 90% of the computation, and several other vector operations (additions/subtractions, scalar products, min/max operations). Hence, most of the code optimization effort address the matrix-vector multiplication, originally implemented with two nested loops. In particular, the following actions were undertaken:

- unroll the inner loop, pipeline the outer loop, and suitably partition the matrix and vector,
- speed up the summation in the scalar product corresponding to each matrix row by adopting the binary tree approach,
- in case of a tall matrix, subdivide the matrix vertically.

Table 1 summarizes the FPGA execution times and the resource usage of three code versions. The initial version was compiled for the popular Xilinx Zedboard hardware platform; for the computation of each control signal sample it requires $4.5 \cdot 10^6$ clock periods (clk), which amounts to 45 ms. In optimized version A, the abovementioned optimization steps were undertaken attempting to maximize the usage of the available Zedboard resources for parallelization; the sample computation requires $0.5 \cdot 10^6$ clk, amounting to 5 ms which is still longer than the computation using a CPU. Choosing a larger FPGA ZC706, with which a higher degree of parallelization is achieved, the sample computation requires $0.12 \cdot 10^6$ clk or 1.2 ms, which finally means a substantial acceleration compared to the initial CPU implementation.

TABLE 1
PERFORMANCE AND RESOURCE SUMMARY TABLE

|  | *Initial version* | *Optimized version A* | *Optimized version B* |
|---|---|---|---|
| *Execution time:* | | | |
| Single iteration latency | ~9000 clk | ~1000 clk | 237 clk |
| Sample computation latency | $4.5 \cdot 10^6$ clk (45 ms) | $0.5 \cdot 10^6$ clk (5 ms) | $0.12 \cdot 10^6$ clk (1.2 ms) |
| *FPGA Resource usage:* | Zedboard | Zedboard | ZC706 |
| BRAM | 87 (31%) | 251 (90%) | 533 (50%) |
| DSP | 27 (12%) | 140 (64%) | 675 (75%) |
| Flip-Flop | 18088 (17%) | 70209 (66%) | 122085 (27%) |
| Look-Up Tables | 12768 (24%) | 53925 (101%) | 150765 (68%) |

## III. CONCLUSIONS

With the FPGA implementation using the Xilinx ZC 706 FPGA, the time required for MPC on-line optimization in each sample step of the control algorithm was shortened to 1.2 ms, which a third of the initial CPU implementation.

With respect to the anticipated sample period of 100 ms for the PCSC controller in Iter, the computation time is considered sufficiently short for the implementation. However, for testing and evaluation of MPC control on dynamically faster medium-sized tokamaks, even faster implementation is desired. It is expected that a substantial further acceleration of execution is possible by using fixed-point or integer arithmetics. Unfortunately, this will even increase the already undesirably high extent of manual modifications of the code; this could be avoided by using FPGAs featuring hardware floating-point arithmetic operations, but is questionable from the cost-effectiveness point of view. Further acceleration may also be achieved by using a larger FPGA with more resources.

**Samo Gerkšič** completed his studies of electrical engineering with PhD at the Faculty of Electrical Engineering,

University of Ljubljana in 2000.

Currently he is a Senior R&D associate at JSI E2. His research interests include advanced control algorithms, modelling, identification and simulation of dynamical systems and in control applications. Recently he was most active in the application of advanced model predictive controllers to plasma magnetic control for tokamak fusion reactors as the principal investigator of the EUROfusion project FMPCFMPC "Fast Model Predictive Control for Magnetic Plasma Control". He is the author of more than 15 scientific articles and more than 40 conference publications.

**Boštjan Pregelj** completed his studies with Ph.D. in Electrical Engineering at the Faculty of Electrical Engineering, University of Ljubljana in 2009.

He is a Research associate at JSI E2. His research interests include modeling and simulation, predictive control, supervisory control and hybrid systems. Since 2009, his research has been predominantly oriented towards fuel cells and reformer systems applications. Recently, he has also been working also on the development of fast MPC for Iter plasma control. He has been involved in a number of national, EU FP, and Horizon Research Projects, recently, also as the coordinator of the FP7 FCGEN Project. He is the author of 14 scientific articles and 40 conference publications.

**Matija Perne** graduated at the Faculty of Mathematics and Physics, University of Ljubljana, in 2007 and finished his Ph.D. at the Graduate School of the University of Nova Gorica in 2012.

Formerly he was employed at Domel d.d. and at the Karst Research Institute of the Research Centre of the Slovenian Academy of Sciences and Arts. Since 2012 he is with the Department of Systems and Control at Jožef Stefan Institute, Ljubljana, Slovenia, where he is a Scientific associate. His research interests include mathematical modelling of speleogenesis and fuel cell processes, and optimization techniques.